\documentclass{emulateapj}

\newcommand{\htwo}{\mbox{\rm H$_2$}}

\newcommand{\halpha}{\mbox{\rm H$\alpha$}}

\newcommand{\xco}{\mbox{$X_{\rm CO}$}}
\newcommand{\xcounits}{\mbox{cm$^{-2}$ (K km s$^{-1}$)$^{-1}$}}
\newcommand{\sigsfr}{\mbox{$\Sigma_{\rm SFR}$}}

\newcommand{\sightwo}{\mbox{$\Sigma_{\rm H2}$}}

\shorttitle{A Constant Molecular Gas Depletion Time in Nearby Disk Galaxies}
\shortauthors{F.~Bigiel et al.}

\begin{document}

\slugcomment{Accepted for Publication in ApJL}
\title{A Constant Molecular Gas Depletion Time in Nearby Disk Galaxies}

\author{F.~Bigiel\altaffilmark{1}, A.K.~Leroy\altaffilmark{2,10}, F.~Walter\altaffilmark{3}, E.~Brinks\altaffilmark{4},
W.J.G.~de~Blok\altaffilmark{5}, C.~Kramer\altaffilmark{6}, H.W.~Rix\altaffilmark{3}, A.~Schruba\altaffilmark{3}, K.-F.~Schuster\altaffilmark{7},
A.~Usero\altaffilmark{8}, H.W.~Wiesemeyer\altaffilmark{9}}

\altaffiltext{1}{Department of Astronomy, Radio Astronomy Laboratory,
University of California, Berkeley, CA 94720, USA; bigiel@astro.berkeley.edu}
\altaffiltext{2}{National Radio Astronomy Observatory, 520 Edgemont Road, Charlottesville,
VA 22903, USA}
\altaffiltext{3}{Max-Planck-Institut f{\"u}r Astronomie, K{\"o}nigstuhl 17,
69117 Heidelberg, Germany}
\altaffiltext{4}{Centre for Astrophysics Research, University of
Hertfordshire, Hatfield AL10 9AB, UK}
\altaffiltext{5}{Astrophysics, Cosmology and Gravity Centre, Department of Astronomy,
University of Cape Town, Private Bag X3, Rondebosch 7701, South Africa}
\altaffiltext{6}{IRAM, Avenida Divina Pastora 7, E-18012 Granada, Spain}
\altaffiltext{7}{IRAM, 300 rue de la Piscine, 38406 St. Martin d'H{\`e}res, France}
\altaffiltext{8}{Observatorio Astron{\'o}mico Nacional, C/ Alfonso XII, 3, 28014, Madrid, Spain}
\altaffiltext{9}{Max-Planck-Institut f{\"u}r Radioastronomie, Auf dem H{\"u}gel 69,
53121 Bonn, Germany} 
\altaffiltext{10}{Hubble Fellow}

\begin{abstract}
We combine new sensitive, wide-field CO data from the HERACLES survey
with ultraviolet and infrared data from GALEX and {\em Spitzer} to
compare the surface densities of \htwo , $\Sigma_{\rm H2}$, and the
recent star formation rate, $\Sigma_{\rm SFR}$, over many thousands of
positions in 30 nearby disk galaxies. We more than quadruple the
  size of the galaxy sample compared to previous work and include
  targets with a wide range of galaxy properties. Even though the
disk galaxies in this study span a wide range of properties, we find a
strong, and approximately linear correlation between $\Sigma_{\rm
  SFR}$ and $\Sigma_{\rm H2}$ at our common resolution of 1~kpc. This
implies a roughly constant median H$_2$ consumption time, $\tau_{\rm
  Dep}^{\rm H2} = \Sigma_{\rm H2} / \Sigma_{\rm SFR}$, of
$\sim2.35$\,Gyr (including heavy elements) across our sample. At 1~kpc
resolution, there is only a weak correlation between $\Sigma_{\rm H2}$
and $\tau_{\rm Dep}^{\rm H2}$ over the range $\Sigma_{\rm H2} \approx
5$--$100$~M$_\odot$~pc$^{-2}$, which is probed by our data. We compile
a broad set of literature measurements that have been obtained using a
variety of star formation tracers, sampling schemes and physical
scales and show that overall, these data yield almost exactly the same
results, although with more scatter. We interpret these results as
strong, albeit indirect evidence that star formation proceeds in a
uniform way in giant molecular clouds in the disks of spiral galaxies.
\end{abstract}


\section{Introduction}
\label{sec:intro}

Giant molecular clouds (GMCs) are the sites of star formation in the Milky Way
\citep[e.g.,][]{BLITZ93}. Therefore, it is not surprising that a
strong correlation is observed between tracers of molecular gas and
recent star formation
\citep[e.g.,][]{ROWND99,WONG02,LEROY08,BIGIEL08}, while the
correlation between atomic gas and recent star formation is found to be weak or
absent within galaxies \citep[e.g.,][]{KENNICUTT07, BIGIEL08}. The details of this
correlation have important implications. Its evolution over cosmic
time informs our understanding of galaxy assembly
\citep{DADDI10,GENZEL10}. The finding of short molecular gas
consumption times compared to galaxy lifetimes highlights the
importance of fueling the inner disks of galaxies. The relatively low
efficiency of star formation per dynamical time requires that the star
formation process be more complex than simple gravitational collapse
\citep[e.g.,][]{MCKEE07}. Finally, the relationship between star
formation and molecular gas is an important input and benchmark for
models attempting to reproduce today's galaxies or galaxy populations.

The importance of this topic has led to several studies of the
relationship between surface densities of H$_2$ and the star formation
rate. Many of these focus on single galaxies
\citep[e.g.,][]{HEYER04,KENNICUTT07,SCHUSTER07,BLANC09,VERLEY10,RAHMAN10} or a
small sample
\citep[e.g.,][]{WONG02,LEROY08,BIGIEL08,WILSON09,WARREN10}. Restricted
by the availability of sensitive and wide-field molecular gas maps,
studies of large sets of galaxies
\citep[e.g.,][]{YOUNG96,KENNICUTT98A,ROWND99,MURGIA02,LEROY05} mostly
used integrated measurements or a few pointings per galaxy.  To date,
no homogeneous analysis of the correlation between the star formation
rate and H$_2$ surface densities in a large set of nearby galaxies at
good spatial resolution exists.

In this letter we take this next logical step, comparing molecular gas
--- traced by CO emission --- to recent star formation --- traced by
ultraviolet and infrared emission --- at $1$~kpc resolution
across a large sample of 30 nearby galaxies. This sample is
  significantly larger and more diverse than that of
  \citet[][hereafter B08]{BIGIEL08}. From 2007-2010, the HERA CO-Line
Extragalactic Survey \citep[HERACLES, first maps are presented
  in][]{LEROY09} collaboration used the IRAM 30-m
telescope\footnote{IRAM is supported by CNRS/INSU (France), the MPG
  (Germany) and the IGN (Spain).}  to construct maps of CO
$J=2\rightarrow1$ emission from 48 nearby galaxies. Because the
targets overlap surveys by {\em Spitzer} \citep[mostly
  SINGS,][]{KENNICUTT03} and GALEX \citep[mostly the
  NGS,][]{GILDEPAZ07}, excellent multiwavelength data are available
for most targets.

\section{Method}
\label{sec:method}

We study all galaxies meeting the following criteria: 1) a HERACLES
map containing a robust CO $J=2\rightarrow1$ detection, 2) GALEX far
UV (FUV) and {\em Spitzer} infrared data at 24$\mu$m (IR), and 3) an
inclination $\lesssim 75\degr$. The first condition excludes low mass
galaxies without CO detections. The second removes a few targets with
poor {\em Spitzer} 24$\mu$m data. The third disqualifies a handful of
edge-on galaxies. We are left with $30$ disk galaxies, listed in Table
\ref{table1} along with distances adopted from
\citet{walter08}, LEDA, and NED. This sample is more than four times larger
than that of B08 and spans a substantial range in metallicities ($8.36\lesssim z\lesssim8.93$)\footnote{Metallicities are adopted from \citet{moustakas10} where available and supplemented by data from the compilations in \citet{calzetti10} and \citet{marble10}.} and mass ($8.9\lesssim {\rm log}(M_{*})\lesssim 11.0$)\footnote{Stellar masses are estimated using the near IR luminosities from \citet{dale07,dale09} and the mass-to-light ratio from \citet{LEROY08}.}.

\begin{deluxetable}{llll}
\tablecaption{Galaxy Sample} \tablehead{
\colhead{Galaxy} & \colhead{$D$} & \colhead{Galaxy} & \colhead{$D$} \\
\colhead{} & \colhead{[Mpc]} & \colhead{} & \colhead{[Mpc]}} \startdata
  NGC\,0337\tablenotemark{d} & 24.7& NGC\,4254\tablenotemark{d} & 20.0 \\
  NGC\,0628\tablenotemark{B08} & 7.3 &NGC\,4321 & 14.3 \\
  NGC\,0925 & 9.2 &NGC\,4536 & 14.5 \\
  NGC\,2403 & 3.2  &NGC\,4559 & 7.0\\
  NGC\,2841 & 14.1 & NGC\,4569\tablenotemark{d} & 20.0\\
  NGC\,2903 & 8.9  &NGC\,4579\tablenotemark{d} & 20.6\\
  NGC\,2976 & 3.6  &NGC\,4625 & 9.5\\
  NGC\,3049 & 8.9 &NGC\,4725 & 9.3 \\
  NGC\,3184\tablenotemark{B08} & 11.1 &NGC\,4736\tablenotemark{B08} & 4.7\\
  NGC\,3198 & 13.8 &NGC\,5055\tablenotemark{B08} & 10.1 \\
  NGC\,3351 & 10.1 &NGC\,5194\tablenotemark{B08} & 8.0 \\
  NGC\,3521\tablenotemark{B08} & 10.7 & NGC\,5457 & 7.4\\
  NGC\,3627 & 9.3  &NGC\,5713\tablenotemark{d} & 26.5\\
  NGC\,3938 & 12.2 &NGC\,6946\tablenotemark{B08} & 5.9 \\
  NGC\,4214 & 2.9  &NGC\,7331 & 14.7
\enddata
\tablenotetext{B08}{Target from B08.}
\tablenotetext{d}{Too distant to reach 1\,kpc resolution, included in the 1\,kpc plots at their respective native resolution.}
\label{table1}
\end{deluxetable}

We follow the approach of B08 with only
a few modifications. B08 compared the first seven HERACLES maps to FUV,
IR, and H$\alpha$ emission to infer the relationship between the
surface density of H$_2$, $\Sigma_{\rm H2}$, and the star formation
rate surface density, $\Sigma_{\rm SFR}$. As in B08, we estimate
$\Sigma_{\rm H2}$ from HERACLES CO $J=2\rightarrow1$ emission. We
assume a Galactic $\xco = 2 \times 10^{20}$~\xcounits , correct
for inclination, include helium
in our quoted surface densities (a factor of 1.36, a difference from
B08), and adopt a CO line ratio $I (2-1) / I (1-0) = 0.7$ (note that B08
used a ratio of 0.8).

We estimate $\Sigma_{\rm SFR}$ (inclination corrected) using a combination of FUV emission
and $24\mu$m emission. FUV emission traces mainly photospheric emission from O and B stars,
with a typical age of $\sim 20$--$30$~Myr \citep{LEITHERER99,SALIM07}
but sensitive to populations up to 100\,Myr of age. Infrared emission at
24$\mu$m comes from dust mainly heated by young stars. This emission
correlates closely with other signatures of recent star formation,
especially H$\alpha$ emission, and so has been used to correct optical
and UV tracers for the effects of extinction
\citep{CALZETTI07,KENNICUTT07}. \citet{LEROY08} motivated this FUV--IR
combination, showing that it reproduces other estimates of
$\Sigma_{\rm SFR}$ with $\sim 50\%$ accuracy down to $\Sigma_{\rm SFR}
\approx 10^{-3}$~M$_\odot$~yr$^{-1}$~kpc$^{-2}$.

For 24 galaxies, we use FUV maps from the Nearby Galaxy Survey
\citep[NGS,][]{GILDEPAZ07}, for five targets from the
All-sky Imaging Survey (AIS) and for one galaxy we use a map from the
Medium Imaging Survey (MIS). We use maps of IR emission at 24$\mu$m
from the {\em Spitzer} Infrared Nearby Galaxies Survey
\citep[SINGS,][]{KENNICUTT03} and the Local Volume Legacy Survey
\citep[LVL,][]{dale09}. Handling of the maps follows B08.

We convolve the IR and FUV maps to the $13\arcsec$ (FWHM) resolution of
the HERACLES data. Given the wide distance range of our sample,
$13\arcsec$ resolution corresponds to physical scales from $180$~pc to
$1.7$~kpc. To avoid being influenced by physical resolution, we
create a second set of maps at a common physical resolution of 1~kpc
(FWHM), appropriate to carry out a uniform analysis. Five galaxies are
too distant to reach 1~kpc resolution. We include them in our ``kpc''
analysis at their native resolution, 1.4\,kpc on average (excluding
them does not change our conclusions).

The HERACLES maps are masked to include only significant emission \citep{LEROY09}. The
exact completeness of each map in mass surface density depends on the
inclination and, for fixed spatial resolution, the distance of the
target. A typical noise level is $25$~mK per 5.2~km~s$^{-1}$ channel
at $13\arcsec$ resolution. For the most distant, face-on systems this
limitis $I_{\rm CO} > 0.8$~K~km~s$^{-1}$ or 5~M$_\odot$~pc$^{-2}$ for our adopted \xco\ and
line ratio. Closer or more inclined systems will be complete to lower
$\Sigma_{\rm H2}$.

We sample both sets of maps, one at $13\arcsec$ and one at 1\,kpc
resolution, using a hexagonal grid spaced by one half-resolution
element. We keep only sampling points inside the $B$-band 25$^{\rm
  th}$ magnitude isophotal radius, $r_{25}$, and where the HERACLES
mask includes emission. At $13\arcsec$ resolution, this yields
$\Sigma_{\rm SFR}$ and $\Sigma_{\rm H2}$ estimates for a total of $\sim 27,000$
points ($\sim 5,000$ independent measurements) in 30 nearby
star-forming galaxies.  At $1$~kpc resolution, this number drops to
$\sim 12,000$ ($\sim 2000$ independent) measurements.

\section{Results}
\label{sec:results}

\begin{figure*}
\plotone{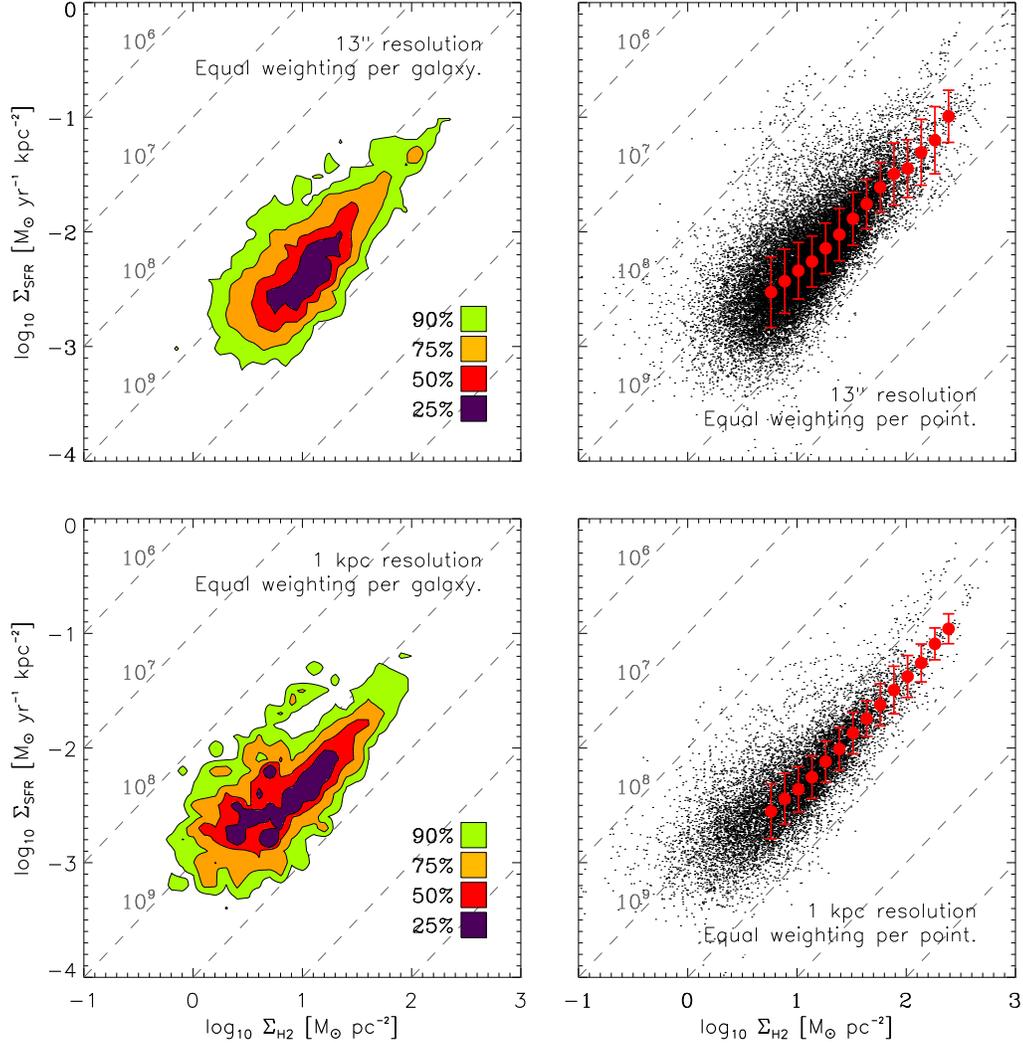}
\caption{Star formation rate surface density, \sigsfr , estimated from
  FUV+24$\mu$m emission as a function of molecular gas surface
  density, \sightwo , estimated from CO $J=2\rightarrow1$ emission for
  30 nearby disk galaxies. The left panels show data density with
  equal weight given to each galaxy. Purple, red, orange, and green
  contours encompass the densest 25, 50, 75, and 90\% of the data. The
  right panels show each measurement individually as a black dot. The
  red points indicate running medians in \sigsfr\ as a function of
  \sightwo\ and the error bars show the 1$\sigma$ log-scatter in each
  \sightwo\ bin. In both panels, dotted lines indicate fixed
  \htwo\ depletion times in yr. Measurements in the top panels are on
  a common angular scale of $13\arcsec$, those in the bottom panels
  are on a common physical scale of 1\,kpc. All panels show a strong
  correlation between \sigsfr\ and \sightwo\ with the majority of data
  having $\tau_{\rm Dep}^{\rm H2} \sim 2.3$\,Gyr.}
\label{fig:combined}
\end{figure*}

\begin{figure}[!t]
\epsscale{1}
\plotone{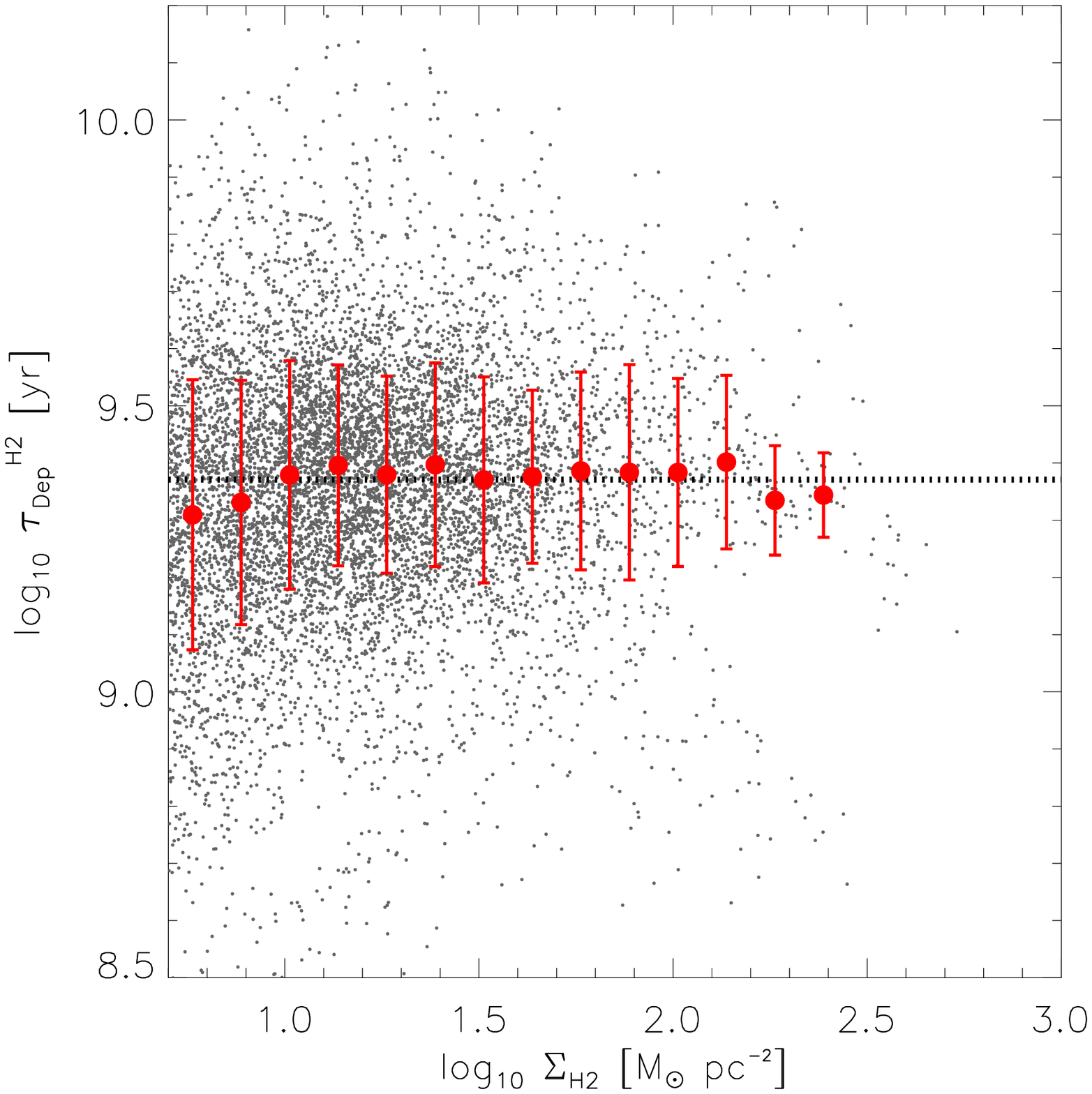}
\caption{$\tau_{\rm Dep}^{\rm H2}$ as a function of \sightwo. Gray
  points indicate individual measurements. Red points show the running
  median. Error bars indicate the 1$\sigma$ scatter in each bin. The
  dashed line shows the median $\tau_{\rm Dep}^{\rm
    H2}\approx2.35$\,Gyr. $\tau_{\rm Dep}^{\rm H2}$ shows little
  or no systematic variation with \htwo\ surface density over the range
  $\Sigma_{\rm H2} \sim 5$--$100$~M$_\odot$~pc$^{-2}$.}
\label{fig:tau}
\end{figure}

Figure \ref{fig:combined} shows our data in
\sigsfr-\sightwo\ space. The upper panels present measurements at a
common angular resolution of $13\arcsec$, the lower panels show
results for a common physical scale of 1\,kpc. The left panels show
contours indicating the density of data with each galaxy weighted
equally. The right panels directly show each data point. Dotted lines
in each plot indicate constant molecular gas depletion times,
$\tau_{\rm Dep}^{\rm H2} = \sightwo/\sigsfr$, i.e., fixed ratios of
H$_2$-to-SFR.

To make the contour plots, we divide the \sigsfr-\sightwo\ space into
0.1\,dex-wide cells to grid the data. During gridding, we assign each
data point a weight inversely proportional to the number of data points for
the galaxy that it was drawn from. This assigns the same total weight
to each galaxy, ensuring that a few large galaxies do not drive the
overall distribution. Contours indicate the density of sampling points
in each cell.

The scatter plots on the right treat all measurements equally, which
leads large galaxies to dominate the distribution. While the contour
plots treat a galaxy as the fundamental unit, the scatter plots treat
each region as a key independent measurement. The red points show a
running median in $\Sigma_{\rm SFR}$ as a function of $\Sigma_{\rm
  H2}$. Though treating $\Sigma_{\rm H2}$ as an independent variable is
not rigorous, this binning is a useful way to guide the eye. We
only bin where $\Sigma_{\rm H2} > 5$~M$_\odot$~pc$^{-2}$ and we are
confident of being complete.

All four plots reveal a strong correlation between \sigsfr\ and
\sightwo . In this letter we focus our quantitative analysis on the
right hand plots, which weight every measurement equally. The Spearman
rank correlation coefficient across all data is $r=0.8$ at 1\,kpc
resolution, indicating a strong correlation between \sigsfr\ and
\sightwo.  We find a median H$_2$ depletion time $\tau_{\rm Dep}^{\rm
  H2} = 2.35$~Gyr with $1\sigma$ scatter 0.24~dex ($\approx
75\%$). The results at fixed $13\arcsec$ resolution are
  similar, median $\tau_{\rm dep}^{\rm H2}$ is $\sim 2.37$~Gyr and
  $r = 0.7$.

It is common to parameterize relationships between gas and star
formation using power law fits. This can be problematic physically,
because data from widely varying environments are often not
well-described by a single power law \citep[B08,][]{bigiel10b}. It is also challenging
practically, because of, e.g., issues of completeness and
upper limits \citep[see][]{BLANC09}, zero point uncertainties \citep[compare][]{RAHMAN10}
or a correct treatment of the uncertainties associated with physical
parameter estimation. Bearing these caveats
  in mind, a rough parameterization may still be useful to the reader.
  If we apply a simple linear regression in log space and fit\footnote{We
    normalize the fit at $\Sigma_{\rm H2} = 10$~M$_\odot$~pc$^{-2}$
    following B08.} the
  function $\Sigma_{\rm SFR} = A \times \left( \Sigma_{\rm H2} / {\rm
    10 M}_\odot~{\rm pc}^{-2}\right)^N$ to the binned kpc data (red points in the lower right
  panel of Figure \ref{fig:combined}), we find $A\approx4.4 \times
  10^{-3}$~M$_\odot$~yr$^{-1}$~kpc$^{-2}$ and $N\approx1.0$. This is
  not rigorous: we have treated the observable $\Sigma_{\rm H2}$ as an
  independent variable and we discarded information in the process of
  binning. However the fit does reasonably bisect the data. We find
  similar results fitting the individual measurements where we are complete with
  $N$ varying by $\pm 0.2$ and $A$ varying by $\sim 30\%$,
  depending mainly on how the fit is constructed.

The results of this fitting can be distilled to what is immediately
apparent from the plot: a characteristic $\tau_{\rm Dep}^{\rm H2} \sim
2.3$~Gyr and a power law index close to unity, so that the data extend
parallel to the dashed lines of fixed $\tau_{\rm Dep}^{\rm H2}$ in
Figure \ref{fig:combined}. The global index close to unity implies
that the ratio of $\Sigma_{\rm H2}$ to $\Sigma_{\rm SFR}$ does not
change much as a function of $\Sigma_{\rm H2}$ across our data. We
quantify this by comparing $\tau_{\rm Dep}^{\rm H2}$ to $\Sigma_{\rm
  H2}$ where we are complete ($\Sigma_{\rm H2} >
5$~M$_\odot$~pc$^{-2}$). Figure \ref{fig:tau} plots the individual
measurements along with a running median and scatter; both show little
or no systematic variation of $\tau_{\rm Dep}^{\rm H2}$ as a function of
$\Sigma_{\rm H2}$ across the range studied. The rank correlation
coefficient relating $\tau_{\rm Dep}^{\rm H2} = \Sigma_{\rm H2} /
\Sigma_{\rm SFR}$ to $\Sigma_{\rm H2}$ is $r = 0.09 \pm 0.01$ in our
kpc data, i.e., the two quantities are only very weakly correlated.

These results extend those found by B08 and \citet{LEROY08}, who
  also found a roughly constant ratio $\Sigma_{\rm H2} /
\Sigma_{\rm SFR}$ for a smaller,
  less diverse sample. Based on detailed studies of Local Group galaxies
  \citep[e.g.][]{BLITZ07,BOLATTO08,BIGIEL10a,FUKUI10}, they speculated that the approximately linear
  \sigsfr-\sightwo\ relation arises because star formation in disk
  galaxies takes place in a relatively uniform population of
  GMCs. Given typical GMC masses of $\sim 10^5$--$10^6$~M$_\odot$ and sizes
  of $\sim 50$~pc, each of our resolution elements likely
  averages over at least a few --- and often many --- GMCs. 
  Thus, in this scenario the relationship
  between $\Sigma_{\rm H2}$ and $\Sigma_{\rm SFR}$ reduces to a
  counting exercise: $\Sigma_{\rm H2}$ corresponds to a different number of GMCs
  inside different resolution elements, rather than to changing physical
  conditions in the molecular gas. This also naturally explains
  the weak dependence of our results on spatial scale, which merely determines
  the number of GMCs per resolution element but leaves the average fixed
  constant $\tau_{\rm Dep}^{\rm H2}$
  intact (compare B08 for a detailed discussion).

This scenario does not contradict earlier results finding that
  $\tau_{\rm Dep}^{\rm H2}$ depends on $\Sigma_{\rm H2}$ or $M_{\rm
    H2}$: the strongest measurements of variable $\tau_{\rm
    Dep}^{\rm H2}$ come from LIRGs and ULIRGs
  \citep[e.g.,][]{KENNICUTT98A,gao04}, systems with \htwo\ surface densities
  significantly exceeding those studied here and where the assumption of a
  uniform GMC population likely breaks down. Departures are
  also expected on scales of individual molecular
  clouds, where only a small fraction of the molecular gas actively forms stars
  \citep[e.g.,][]{Heiderman10}. We will show in the next section, however, that for
  normal disk galaxies and on scales greater than a few 100\,pc  
   our results agree remarkably well with previous measurements of
  $\tau_{\rm Dep}^{\rm H2}$.

\section{Comparison to Literature Data}
\label{sec:lit}

\begin{figure*}[]
\epsscale{1.1} \plottwo{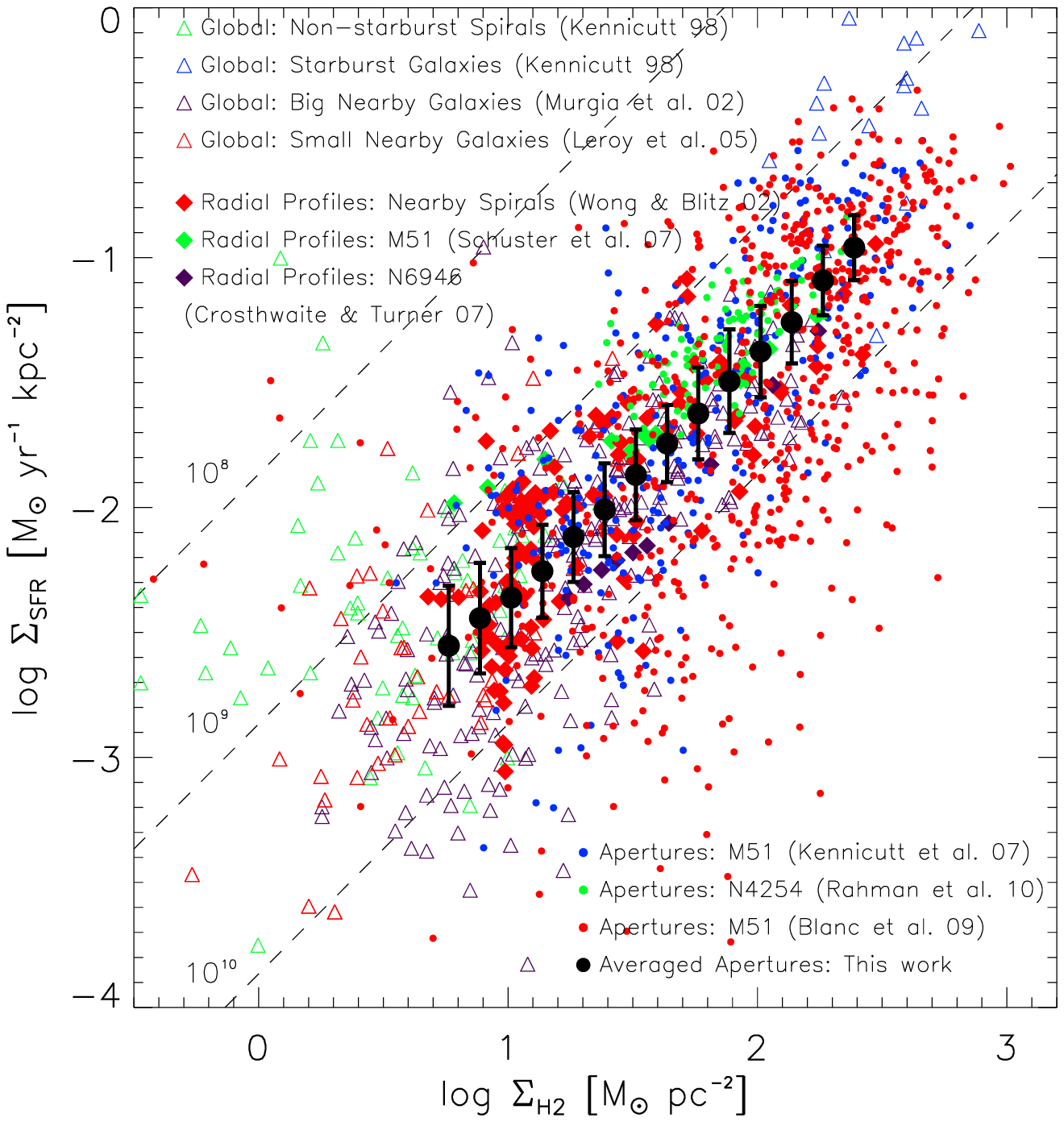}{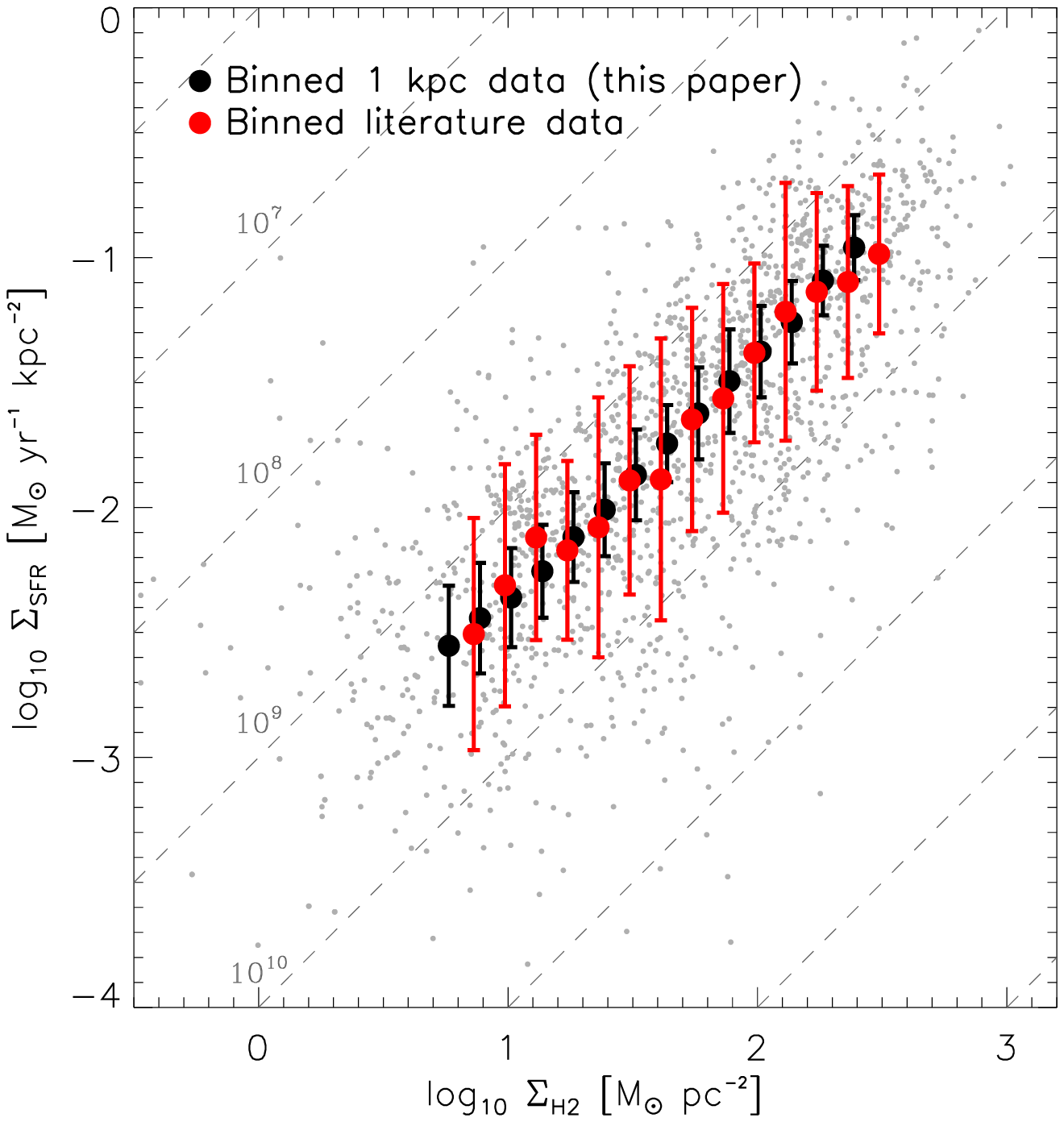}
\caption{\sigsfr\ versus \sightwo\ for a compilation of literature
  measurements and our binned data. In the left panel we label
  individual studies, which employ a wide range of star formation
  tracers, sampling schemes and physical scales. The black points
  indicate the running medians for our 1\,kpc data from Figure
  \ref{fig:combined}. In the right panel we treat all literature
  measurements equally (gray points) and construct a running median
  (red points) in the same way that we binned our data (black points)
  in Figure \ref{fig:combined}. Both panels show excellent agreement
  between our measurements and the literature data and suggest an
  emerging consensus on the basic, approximately linear, $\Sigma_{\rm
    SFR}-\Sigma_{\rm H2}$ relation in nearby disk galaxies.}
\label{fig:lit}
\end{figure*}

As described in Section \ref{sec:intro}, many studies have examined
the relationship between molecular gas and star formation in nearby
disk galaxies over the last decade. The emphasis on power law fits has
somewhat obscured the basic question of whether these data
fundamentally agree or disagree regarding which part of $\Sigma_{\rm
  H2}-\Sigma_{\rm SFR}$ space is occupied by local disk galaxies. To
address this point, Figure \ref{fig:lit} shows our binned data (big
points with error bars) along with a wide compilation of recent
measurements.

We adjust all literature measurements to share our adopted \xco\ and
stellar IMF (Kroupa), but otherwise leave the data unchanged. These
points therefore reflect a wide range of star formation tracers,
sampling schemes and physical scales.

We plot averages over whole galaxies as triangles. These include 57
normal spiral galaxies (green) and 15 starburst galaxies (blue) from
\citet{KENNICUTT98A}. \citet{KENNICUTT98A} estimates $\Sigma_{\rm
  SFR}$ from H$\alpha$ for normal spirals and IR emission for
starbursts. We also show 236 pointings towards spirals from
\citet[][purple]{MURGIA02} and towards 80 small nearby spirals and dwarfs from
\citet[][red]{LEROY05}\footnote{These studies compile measurements from
  \citet{YOUNG95}, \citet{TAYLOR98}, \citet{ELFHAG96}, and
  \citet{BOKER03}.}. Both data sets have $\sim 50\arcsec$ resolution,
corresponding to $\sim1$--$4$~kpc, and use 1.4 GHz radio continuum
(RC) emission to estimate $\Sigma_{\rm SFR}$.

Filled diamonds indicate radial profile measurements. Data for 7
nearby spirals from \citet{WONG02} are shown in red, those for M51
from \citet{SCHUSTER07} in green and those for NGC\,6946 from
\citet{CROSTHWAITE07} in purple. \citet{WONG02} derive \sigsfr\ from
\halpha\ emission, \citet{SCHUSTER07} from RC emission, and
\citet{CROSTHWAITE07} from FIR emission.

Small points represent aperture data. Blue points show 520\,pc-sized
aperture measurements from \citet{KENNICUTT07} of star forming regions
in the spiral arms of NGC\,5194 (M51). They infer $\Sigma_{\rm SFR}$
from a combination of \halpha\ and 24\,\micron\ emission. Green points
show 500\,pc apertures from \citet{RAHMAN10}, who sample mainly the
spiral arms of NGC\,4254. The points shown here reflect \sigsfr\ as
derived from FUV and 24\,\micron\ emission. Red points indicate
170\,pc apertures covering the central $4.1\times4.1$\,kpc$^{2}$ of
NGC\,5194 (M51) from \citet{BLANC09}. They infer \sigsfr\ from extinction
corrected \halpha\ emission using integral field unit
observations. The left panel of Figure \ref{fig:lit} labels these
various studies and overplots our data.

Figure \ref{fig:lit} shows that these measurements sweep out a
distinct part of $\Sigma_{\rm SFR}$--$\Sigma_{\rm H2}$ space. Most
data scatter between $\tau_{\rm Dep}^{\rm H2} = 10^9$ and $10^{10}$~yr
and our measurements lie near the center of the distribution. The
right panel in Figure \ref{fig:lit} shows this most clearly: we take
the simplistic approach of treating all of the literature data equally
(shown as gray points) and construct the same running median that we
use on our own data. The literature average (red points) agrees
strikingly well with our measurements (black points). This implies
  that our results are robust with respect to the choice of tracers or
  experimental setup. The literature sample as a whole also suggests
that $\tau_{\rm Dep}^{\rm H2} \approx 2.3$~Gyr in nearby disks and
that $\tau_{\rm Dep}^{\rm H2}$ is a fairly weak function of
$\Sigma_{\rm H2}$.

\section{Summary}
\label{sec:summary}

Using new IRAM 30m CO $J=2\rightarrow1$ maps from the HERACLES survey,
we determine the relation between \htwo\ surface density, \sightwo,
and SFR surface density, \sigsfr, in 30 nearby disk galaxies.
  This significantly extends the number of galaxies (by more than a
  factor of four) and the range of galaxy properties probed compared
  to \citet{BIGIEL08}. We present our main results for a common physical
resolution of 1\,kpc. We find a remarkably constant molecular gas
consumption time $\tau_{\rm Dep}^{\rm H2} \approx 2.35$~Gyr (including
helium) with a 1$\sigma$ scatter of 0.24\,dex ($\approx 75\%$) and
little dependence of $\tau_{\rm Dep}^{\rm H2}$ on $\Sigma_{\rm H2}$
over the range $\Sigma_{\rm H2} \sim 5$--$100$~M$_\odot$~pc$^{-2}$.

This extends and reinforces the conclusions of \citet{BIGIEL08} and \citet{LEROY08}
that the star formation rate per unit H$_2$ in the disks of massive
star-forming galaxies is, to first order, constant. We interpret this
as strong, yet indirect, evidence that the disks of nearby
spiral galaxies are populated by GMCs forming stars in a relatively
uniform manner. We caution that these results are specific to
  disk galaxies and scales on which we average over many GMCs --- they
  may be expected to break down at very high surface densities and
 on small scales. Taken as a whole, a broad compilation of literature
data on disk galaxies from the last decade yields impressively similar
results.

\acknowledgments We thank the GALEX NGS, SINGS, and LVL teams for
making their outstanding datasets available. We thank Karl Gordon for
the kernel used on the MIPS 24$\mu$m data and Nurur Rahman for sharing
his data. We thank the staff of the IRAM 30m telescope
for their assistance carrying out the survey. F.B., A.K.L., and
F.W. gratefully acknowledge the Aspen Center for Physics, where part
of this work was carried out. Support for A.K.L. was provided by NASA through Hubble
Fellowship grant HST-HF-51258.01-A awarded by the Space Telescope
Science Institute, which is operated by the Association of
Universities for Research in Astronomy, Inc., for NASA, under contract
NAS 5-26555. The work of W.J.G.d.B. is based upon research supported
by the South African Research Chairs Initiative of the Department of Science and
Technology and National Research Foundation.

\end{document}